\documentclass[11pt,a4paper]{article} 
\pdfoutput=1
\usepackage{jheppub}

\usepackage{amsmath, amssymb}
\usepackage{mathpazo}
\usepackage{mathrsfs}
\usepackage{array,arydshln}

\usepackage{graphicx,epsfig}
\usepackage{epic}
\usepackage{color}
\usepackage{youngtab}
\usepackage{float}

\newcommand{\be}{\begin{equation}}
\newcommand{\ee}{\end{equation}}
\newcommand{\ba}{\begin{eqnarray}}
\newcommand{\ea}{\end{eqnarray}}

\newcommand{\mc}{\mathcal }

\newcommand{\der}{\partial}
\newcommand{\derbar}{\overline\partial}
\newcommand{\mb}{\mathbb }

\newcommand{\reni}{R\'enyi }
\newcommand{\ordx}[1]{\mc O(x^{#1})}

\newcommand{\mk}{\mathfrak}

\def\XXint#1#2#3{{\setbox0=\hbox{$#1{#2#3}{\int}$}
     \vcenter{\hbox{$#2#3$}}\kern-.5\wd0}}

    \newcommand{\beq}{\begin{equation}}
    \newcommand{\eeq}{\end{equation}}
    \newcommand\beqa{\begin{eqnarray}}
    \newcommand\eeqa{\end{eqnarray}}

\title{On the next-to-leading holographic entanglement entropy in $AdS_{3}/CFT_{2}$}

\author[a]{Matteo Beccaria} 
\author[a]{, Guido Macorini}

\affiliation[a]{Dipartimento di Matematica e Fisica Ennio De Giorgi,\\
Universit\`a del Salento \& INFN, Via Arnesano, 73100 Lecce, 
Italy} 
                     
\emailAdd{matteo.beccaria@le.infn.it}
\emailAdd{guido.macorini@le.infn.it}

\abstract{
We reconsider the one-loop correction to the holographic entanglement entropy in $AdS_{3}/CFT_{2}$
by analysing the contributions due to a bulk higher spin $s$ current or a scalar field with scaling dimension $\Delta$.
We consider the two-interval case and work perturbatively in their small cross ratio $x$.
We provide various results for the entanglement entropy due to the so-called CDW elements of the associated 
Schottky uniformization group. In particular, in the higher spin current case, we obtain a closed formula 
for all the contributions of the form $\mc O(x^{2s+p})$ up to $\mc O(x^{4s})$, where 2-CDW elements are relevant.
In the scalar field case, we calculate the similar contributions for generic values of $\Delta$. 
The terms up to $\mc O(x^{2\Delta+5})$ are compared with an explicit CFT calculation with full agreement.
The analysis exploits  various 
simplifications which are valid in the strict entanglement limit of the \reni entropy. 
This allows to identify in a clean way the 
relevant operators that provide the gravity result. The 2-CDW contributions are also analysed and a closed formula
for the leading $\ordx{4s}$ coefficient is presented as a function of the generic spin $s$. As a specific application, we 
combine the CDW and 2-CDW calculations and present
the complete $\ordx{4s+2}$ entanglement entropy for a spin $s=2,3,4$ higher spin currents.
}

\allowdisplaybreaks

\begin{document} \maketitle

\bigskip

\section{Introduction and summary of results}

A quite general feature of quantum gravity is the holographic principle. Roughly speaking, it states that 
the number of degrees of freedom of a $(d + 2)$-dimensional quantum gravity theory is 
comparable to  that of a non-gravitational quantum system in 
$d+1$ dimensions \cite{Hooft:1993gx,Susskind:1994vu}. This remarkable feature of gravity theories 
is one of the most physically intriguing aspects of the 
Bekenstein-Hawking formula \cite{Bekenstein:1973ur,Hawking:1974sw} for the entropy $S_{\rm BH}$
of a black hole. Indeed, $S_{\rm BH}$ turns out to be proportional to the area of the black hole event horizon
instead of its volume. In recent times, the powerful framework of AdS/CFT correspondence \cite{Maldacena:1997re}
allowed to study several theoretical laboratories with explicit examples of  theories where
quantum gravity on $(d + 2)$-dimensional anti-de Sitter spacetime $AdS_{d+2}$ is 
holographically equivalent to a  conformal field theory ${\rm CFT}_{d+1}$.

\vskip 5pt
A very interesting open problem in AdS/CFT is to characterise the region of the $AdS$
 bulk space that encodes the information associated with the dual conformal 
 theory \cite{Ryu:2006ef,Takayanagi:2012kg}. From this perspective, a promising quantity to answer this question
is the entanglement entropy. It is a universal (non-local) observable in the sense that its definition does not depend on the 
microscopic details of the theory under consideration. Also, it can be defined in full generality for a
quantum system. Indeed, it is simply the von Neumann entropy $S_{A}$ associated with the reduced density matrix 
obtained by integrating out the degrees of freedom inside a $d$-dimensional space-like subspace $B$, the
complement of $A$, of a given $(d + 1)$- dimensional quantum field theory. 
The entanglement entropy $S_{A}$ measures how the subsystems A and B are correlated with each other
and has some analogy with black hole entropy. The entanglement entropy is expected to 
be related to the number of degrees of freedom. Indeed, in two-dimensional conformal field theory, general arguments show that it is proportional to the central charge  $c$ \cite{Holzhey:1994we}.
Finally, as opposed to the usual thermal entropy, the entanglement entropy  does not vanish at zero temperature
and therefore it can probe the properties of the ground state of a given quantum system.

\vskip 5pt
A remarkable AdS/CFT result for the calculation of the entanglement entropy is the 
celebrated Ryu-Takayanagi formula \cite{Ryu:2006bv,Ryu:2006ef}.
It states that, at leading order in the semiclassical expansion, 
\be
S_{A} = \frac{\mbox{Area}(\Sigma_{A})}{4\,G_{N}^{(d+2)}},
\ee
where $\Sigma_{A}$ is the $d$-dimensional minimal surface whose boundary is  the $(d-1)$-dimensional 
manifold $\partial\Sigma_{A} = \partial A$, the interface between $A$ and $B$, and the constant $G_{N}^{(d+2)}$ is  Newton's constant 
in $AdS_{d+2}$. The Ryu-Takayanagi formula can be tested by  quantum field theoretical calculations in the dual 
conformal theory.
At {leading order} in the $1/c$ expansion \cite{Headrick:2010zt,Hartman:2013mia,Faulkner:2013yia} one finds that 
when $A$ is a set of disjoint intervals with lengths $\ell_{i}$, the entanglement entropy is rigorously 
\be
S_{A} = \frac{c}{3}\sum_{i}\log\frac{\ell_{i}}{a}+\mc O(c^{0}),
\ee
where $a$ is an ultraviolet small length scale. This scaling form is 
in agreement with old calculations \cite{Bombelli:1986rw,Srednicki:1993im,Holzhey:1994we}. Also, it can be shown
that this expression agrees with the Ryu-Takayanagi formula, using the AdS/CFT relation $\ell_{AdS}/G_{N} = 2c/3$, 
see \cite{Headrick:2010zt}.

\vskip 5pt
The actual calculation of $S_{A}$ is done by the replica trick that amounts to the calculation of the 
$n\to 1$ limit of the $n$-th order \reni entropy  \cite{Calabrese:2009ez}
\be
S_{A} = -\lim_{n\to 1}\frac{\partial}{\partial n}\mbox{Tr}_{A}\rho^{n}_{A},
\ee
where $\rho_{A}$ is the reduced density matrix and $A$ is a disjoint union of intervals. 
The trace is computed by evaluating the partition function $Z_{n}$
on a $n$-sheeted {Riemann surface} with branch cuts at the endpoints of $A$ as 
\be
\mbox{Tr}_{A}\rho^{n}_{A} = \frac{Z_{n}}{Z_{1}^{n}}.
\ee
This complicated ratio can be evaluated directly in CFT by introducing suitable twist operators to implement the 
branch cuts. Alternatively, it can be computed by exploiting AdS/CFT.

\vskip 5pt
The analysis of the  subleading corrections $\mc O(c^{0})$ has been initiated in \cite{Barrella:2013wja}.
In the semiclassical expansion of the partition function at large central charge, the leading correction to the on-shell gravity action is expressed in terms of functional determinants of the operators describing quadratic fluctuations of all the bulk fields. This one-loop correction is perturbatively exact in pure three dimensional 
gravity \cite{Maloney:2007ud}. An explicit elegant expression for functional determinants on 
quotients of $AdS_{3}$ by a Schottky group  has been obtained in \cite{Yin:2007gv,Giombi:2008vd}.

\vskip 5pt
Using these tools, the one-loop entanglement entropy for the two-interval case has been computed 
perturbatively at small cross ratio between the four endpoints in \cite{Barrella:2013wja}. The authors of this paper
also considered the case of a single interval on a circle at finite temperature. In particular, they found  that, at high temperatures,  the one-loop contributions introduce  finite size corrections to the entanglement entropy that are not present classically, see also \cite{Datta:2013hba}. Recently,  the AdS/CFT matching of the entanglement entropy has been extensively 
tested in the series of papers \cite{Chen:2013kpa,Chen:2013dxa,Chen:2014kja}.
In particular, the holographic \reni entropy has been  computed in higher spin $AdS_{3}$ gravity with 
additional $\mc W_{3}$ or $\mc W_{4}$ symmetry at order $\ordx{8}$ in the small cross ratio $x$ for the case of two intervals.
The predictions from gravity and CFT perfectly agree and the explicit results for a spin $s$ higher spin current are
\ba
\label{eq:known-results}
S^{(s=2)\,\rm one-loop} &=& -\frac{x^4}{630}-\frac{2 x^5}{693}-\frac{15 x^6}{4004}-\frac{x^7}{234}-\frac{167 x^8}{36036}+\ordx{9}, \nonumber \\
S^{(s=3)\,\rm one-loop} &=&-\frac{x^6}{12012}-\frac{x^7}{4290}-\frac{7 x^8}{16830}+\ordx{9}, \\
S^{(s=4)\,\rm one-loop} &=&-\frac{x^8}{218790}+\ordx{19}.\nonumber
\ea
Similar calculations has been performed at order $\ordx{6}$
in various massive gravities at the critical points.

\vskip 5pt
Recently, in \cite{Perlmutter:2013paa}, the \reni entropy contributions of Virasoro primaries with arbitrary scaling weights has been computed to leading and next-to-leading order in the interval size. When the primaries are higher spin currents, such terms are placed in one-to-one correspondence with terms in the bulk 1-loop determinants for higher spin gauge fields propagating on handlebody geometries. Deformation by chemical potentials associated with higher spin currents
has been very recently addressed in \cite{Datta:2014ska} with promising results. 

\vskip 5pt
An interesting missing piece in this series of checks are bulk scalars. A motivation for their study is in the 
recent discovery that Vasiliev higher spin theory on $AdS_{3}$ \cite{Vasiliev:1995dn,Vasiliev:1996hn}  
coupled to a complex scalar is
dual to a class of coset  WZW conformal theories in certain large-$N$ limit \cite{Gaberdiel:2010pz}.

\vskip 5pt
In this paper, we elaborate on the one-loop corrections to the entanglement entropy. Our main results 
are explicit closed formulae for the  entanglement entropy due to a higher spin current as well as 
accurate series expansions for the contribution from scalars. We work in the two-interval case 
and in the regime where their cross ratio $x$ is a small parturbative parameter. 
Technically, the contributions to the entanglement entropy can be written as sums over Schottky group elements $\gamma$
and organised according to the complexity of $\gamma$. In more details, one can consider specific
classes of elements $\gamma$  known as $N$-CDW elements. For a higher spin $s$ current, the contribution to the entropy from 
each class starts at order $\ordx{2\,N\,s}$. The same holds for a scalar field with its scaling dimension $\Delta$ 
playing the role of  $s$.

\vskip 5pt
We analyse in great detail the CDW contributions ($N=1$) for both the higher spin currents and a scalar field.
In particular, we derive, in the current case, the following exact nice formula 
\ba
&& S^{(s)\, \rm one-loop} = \\
&& \qquad -\frac{\sqrt\pi}{2}\,\left(\frac{x}{4}\right)^{2s}\,\frac{\Gamma(2s+1)}{\Gamma(2s+\frac{3}{2})}\,\,{}_{3}F_{2}\bigg(2s, 2s-\frac{1}{2}, 2s+1; 2s+\frac{3}{2}, 4s-1; x\bigg)+\ordx{4s}.\nonumber
\ea
where we have explicitly indicated that the formula holds for all terms up to $\ordx{4s-1}$. At order $\ordx{4s}$
other CDW contributions modify this formula together with 2-CDW terms. In the paper, we discuss several series expansions for these further contributions and we give them parametrically in $s$ as far as possible. 
One remarkable result is a closed formula for the leading 2-CDW contribution that reads
\be
S^{(s)\,\rm one-loop}_{\rm 2-CDW} = \sigma_{s}\,x^{4s}+\mc O(x^{4s+1}), 
\ee
where the coefficient $\sigma_{s}$ is obtained from 
\be
\label{eq:sigma}
\sigma_{s} =  -\frac{1}{2}\left[
\frac{1}{\Gamma(2s+1)^{4}}\,(p-s)_{2s}^{4}-\frac{1}{\Gamma(4s+1)^{2}}\,(p-2s)_{4s}^{2}
\right]_{p^{N}\to -N\,\zeta(1-N)},
\ee
where $(a)_{b}$ is a Pochammer symbol.

\vskip 5pt
To give an idea of the accuracy allowed by the tools developed in this paper, we improved  the known expansions (\ref{eq:known-results}) by extending them up to the  order
$\ordx{4s+2}$ with the explicit result
\ba
S^{(s=2)\,\rm one-loop} &=& -\frac{x^4}{630}-\frac{2 x^5}{693}-\frac{15 x^6}{4004}-\frac{x^7}{234}-\frac{167 x^8}{36036}-\frac{69422
   x^9}{14549535}-\frac{122 x^{10}}{24871}+\ordx{11}, \nonumber \\
S^{(s=3)\,\rm one-loop} &=&-\frac{x^6}{12012}-\frac{x^7}{4290}-\frac{7 x^8}{16830}-\frac{28 x^9}{46189}-\frac{15 x^{10}}{19019}-\frac{2
   x^{11}}{2093}-\frac{1644627 x^{12}}{1487285800}\nonumber \\
   && -\frac{458893 x^{13}}{371821450}-\frac{224484047
   x^{14}}{166363540200}+\ordx{15}, \\
S^{(s=4)\,\rm one-loop} &=&-\frac{x^8}{218790}-\frac{4 x^9}{230945}-\frac{3 x^{10}}{76076}-\frac{5 x^{11}}{71162}-\frac{11
   x^{12}}{101660}\nonumber \\
   &&-\frac{11 x^{13}}{72675}-\frac{1001 x^{14}}{5058180}
    -\frac{143 x^{15}}{580754}-\frac{253702367
   x^{16}}{859544957700}\nonumber \\
   && -\frac{1550029508 x^{17}}{4512611027925}-\frac{207186247
   x^{18}}{530052723915}+\ordx{19}.\nonumber
\ea
The case of a scalar field is also considered in great detail. In this case, we also present a detailed discussion 
of the matching with a direct CFT calculation. In this way, we reproduce the gravity calculation 
of the entropy at order up to $\ordx{2\Delta+5}$. This comparison is not surprising, but is somewhat interesting in itself
due to various simplifications that we emphasise in the strict entanglement limit $n\to 1$.

\vskip 5pt
The plan of the paper is the following. In Sec.~(\ref{sec:setup}), we briefly recall the technical tools that are
needed in order to compute the holographic entanglement entropy both in gravity and in the conformal dual.
In Sec.~(\ref{sec:gravity}), we present our results for the gravity prediction of the one-loop entanglement entropy
due to a higher spin current or a scalar. For a higher spin $s$ current, we provide  a closed formula for all 
the terms starting from $x^{2s}$ and up to $x^{4s}$. The terms starting from $x^{4s}$ are also computed, but 
they are only one part of the entanglement entropy that gets corrections also from the so-called 2-CDW contributions
which are later. For a scalar field with scaling dimension $\Delta$, we provide similar expansions
starting from $x^{2\Delta}$ for generic values of $\Delta$. 
In Sec.~(\ref{sec:setup}), we present a detailed CFT check 
of the scalar result for the terms up to $\mc O(x^{2\Delta+5})$. Finally, in Sec.~(\ref{sec:2-CDW}), we 
analyse the 2-CDW contributions providing, among other things, an analytic formula for the leading
$\ordx{4s}$ correction to the entanglement entropy of a higher spin $s$ current. 

\section{Setup of the gravity and CFT calculations}
\label{sec:setup}

As we outlined in the Introduction, the ground state \reni  entropy of a quantum system is defined in general terms as follows. 
We split the spatial domain into two subspaces and build the reduced density matrix $\rho$ by taking the trace over one
of the two subspaces. Then, the  \reni entropy, for integer $n>1$ is defined as 
\be
S_{n} = -\frac{1}{n-1}\log\mbox{Tr}(\rho^{n}).
\ee
The  entanglement entropy is the analytic continuation at $n=1$
\be
S = \lim_{n\to 1}S_{n} = -\mbox{Tr}(\rho\log\rho).
\ee
In a 2d dimensional CFT, the above spatial regions are unions of $N$ disjoint intervals with endpoints $\{z_{i}\}_{i=1, \dots, 2\,N}$ and the \reni entropy can be computed on a higher genus Riemann surface $\Sigma$ by the replica trick, or 
in terms of correlation functions of $2\,N$ twist operators at the endpoints.

In the former approach, $\Sigma$ is an $n$-sheeted cover of the original spacetime, with branch cuts 
between the pairs of endpoints of the  disjoint intervals \cite{Nishioka:2009un}. 
In the AdS/CFT correspondence, the partition function of the 2d CFT on $\Sigma$ is equal to the partition function
of the dual gravity theory on a certain quotient $AdS_{3}/\Gamma$ with conformal boundary $\Sigma$.
Here, $\Gamma$ is a discrete subgroup of the isometry group $PSL(2, \mathbb C)$ of $AdS_{3}$
that realizes $\Sigma$ as $\mathbb C/\Gamma$, i.e. a (replica symmetric) Schottky uniformization of $\Sigma$ (see 
\cite{Barrella:2013wja,Faulkner:2013yia} for details).

The calculation of  $S_{n}$ on the gravity side amounts to the calculation of the higher genus partition function
by means of the bulk gravitational path integral over metrics asymptotic to $\Sigma$ with suitable 
boundary conditions~\cite{Faulkner:2013yia}. In Euclidean signature $AdS_{3}$ is $\mathbb H^{3}$ and we look for 
contributions to the bulk path-integral from locally $\mathbb H^{3}$ handlebody solutions $\mc M$ such that on the boundary we reconstruct $\Sigma=\mathbb C/\Gamma$. The path integral is evaluated in a semiclassical expansion and, at leading order $\mc O(c)$
at large central charge, we need the renormalised on-shell bulk action $S_{\rm grav}(\mc M)$ with 
\be
S_{n} = -\frac{1}{n-1}\,S_{\rm grav}(\mc M)+\mc O(c^{0}).
\ee
The leading term is completely determined by the details of the Schottky uniformization procedure
\cite{Hartman:2013mia,Faulkner:2013yia} and we shall not discuss it any more.

Quantum corrections start at $\mc O(c^{0})$ and are sensitive to the microscopic details, {\em i.e.} the specific form 
of the CFT. Again, AdS/CFT provides a recipe for the calculation. One has to evaluate the 1-loop functional determinants
on handle body geometries for the various bulk fields. The general formula is~\cite{Giombi:2008vd}
\be
\label{eq:GravityPartition}
Z_{\mathbb H^{3}/\Gamma}^{\rm one-loop} = 
\prod_{\phi}\prod_{\gamma\in\mc P}\left[Z^{\Phi}_{\mathbb H^{3}/\mathbb Z}(q_{\gamma},\overline q_{\gamma})\right]^{1/2},
\ee
where $\{\Phi\}$ are the bulk fields, $q_{\gamma}$ are the squared eigenvalues of a certain class $\mc P$ of elements of the Schottky group $\Gamma$, to be specified later in full details. The one-loop contribution to the \reni entropy is then
\be
\label{eq:Gravityformula}
S_{n}^{\rm one-loop} = -\frac{1}{n-1}\log Z^{\rm one-loop}_{\mathbb H^{3}/\Gamma}.
\ee
The building blocks, on the torus $\mathbb C/\mathbb Z$ have been computed 
in ~\cite{Giombi:2008vd} for various fields. In particular, for a higher spin gauge field with spin $s$ one has
\be
\label{eq:dethigher}
Z_{\mathbb C/\mathbb Z}^{(s)}(q, \overline q) = \prod_{n=s}^{\infty}\frac{1}{|1-q^{n}|^{2}},
\ee
while for a scalar operator with conformal dimensions
$(h, \overline h) = (\Delta/2, \Delta/2)$, one has
\be
\label{eq:detscalar}
Z_{\mathbb C/\mathbb Z}^{\rm scalar}(q, \overline q) = \prod_{\ell, \ell'=0}^{\infty}\frac{1}{1-q^{\ell+\frac{\Delta}{2}}\,\overline
q^{\ell'+\frac{\Delta}{2}}}.
\ee
The details for the calculation of the partition function (\ref{eq:GravityPartition}), {\em i.e.} the definition
of $\mc P$ and $q_{\gamma}$ can be found in \cite{Barrella:2013wja} and will be reviewed later. Remarkably, 
the eigenvalue $q_{\gamma}$ is independent on $\Phi$ and is universal in the sense that it can be computed in pure gravity.

The CFT side of this story is also straightforward and can be performed by the method of 
\cite{Hartman:2013mia,CC1,CC2} whose notation we follow. We consider the case of two-intervals. 
The trace of the replica density matrix
is a twist fields four point function in the orbifold theory on $(\mb C/\mb Z)^{\otimes n}$
\be
\mbox{Tr}(\rho^{n}) = \langle\Phi_{+}(z_{1})\,\Phi_{-}(z_{2})\,\Phi_{+}(z_{3})\,\Phi_{-}(z_{4})\rangle_{(\mb C/\mb Z)^{\otimes n}}.
\ee
Conformal symmetry is used to fix the positions to $(z_{1},z_{2},z_{3},z_{4}) = (0,\ell,1,1+\ell)$. The calculation is 
done in the limit of small (conformal invariant) cross-ratio $x=\frac{z_{12}z_{34}}{z_{13}z_{24}}=\ell^{2}$.
As is known, the calculation can be organised at small $x$ as a sum over correlators of the $SO(2,1)\times SO(2,1)$ conformal families of the orthogonal quasi primaries $\mc O_{K}$ appearing in the fusion rule
\be
[\Phi_{+}]\times [\Phi_{-}] = [1]+\sum_{K}d_{K}\,[\mc O_{K}].
\ee
Denoting by $\Delta_{K}$ the dimension of $\mc O_{K}$ and introducing $\alpha_{K}=\langle \mc O_{K}|\mc O_{K}\rangle$, the trace can be written as a sum over global conformal blocks \cite{Perlmutter:2013paa}
\be
\mbox{Tr}(\rho^{n}) = x^{-2\,\Delta_{\Phi}}\,\sum_{K}\frac{d^{2}_{K}}{\alpha_{K}}\,x^{\Delta_{K}}\,\left|
{}_{2}F_{1}(h_{K}, h_{K}; 2\,h_{K}; x)\right]^{2}.
\ee
The first factor depending on the twist fields dimension $\Delta_{\Phi}$ gives the classical $\mc O(c)$ contribution to the
trace. The remaining part gives the one-loop contribution to the \reni entropy
\be
\label{eq:CFTformula}
S_{n}^{\rm one-loop} = -\frac{1}{n-1}\log\left[
1+\sum_{K\neq \mathbb I}\frac{d^{2}_{K}}{\alpha_{K}}\,x^{\Delta_{K}}\,\left|
{}_{2}F_{1}(h_{K}, h_{K}; 2\,h_{K}; x)\right]^{2}
\right],
\ee
where we have separated out the vacuum exchange contribution. Finally, the OPE coefficients $d_{K}$ can be computed by
the simple formula \cite{CC1,CC2}
\be
d_{K} = \ell^{-\Delta_{K}}\,\lim_{z\to\infty}|z^{2\,h_{K}}|^{2}\,\langle \mc O_{K}(z, \overline z)\rangle_{\mc R_{n,1}},
\ee
where $\mc R_{n,1}$ is the $n$-cover Riemann surface of the single-interval problem for an interval of length $\ell$ 
on the infinite line. The explicit calculation exploits the conformal mapping 
\be
f(z) = (1-\ell/z)^{1/n},
\ee
that maps $\mc R_{n,1}$ to $\mathbb C$.

As we summarised in the introduction, various checks of AdS/CFT matching have been performed in specific cases
including, {\em i.e.} for various bulk fields \cite{Chen:2013kpa,Chen:2013dxa,Chen:2014kja}. In particular, this 
includes the graviton and a higher spin current with spin 3, and 4.

Here, we want to test AdS/CFT in the calculation of the one-loop \reni entropy associated with a generic 
scalar fields. The matching of (\ref{eq:CFTformula}) and (\ref{eq:Gravityformula}) is not obvious and we want to explore 
which operators are relevant on the CFT side to reproduce the gravity result. In the  limit $n\to 1$ this leads
to various simplifications that we exploit to test the matching at high orders. The general structure of the result for  
a scalar field is 
\be
\label{eq:structure}
S^{\rm (scalar)\  one-loop} = \sum_{r=1}^{\infty}x^{2\,r\,\Delta}\sum_{t=0}^{\infty}S^{(r,t)}(\Delta)\,x^{t}.
\ee
The terms with $r=1$ comes on the gravity side from the so-called CDW elements of the Schottky group. On the CFT side, they are associated with descendants of the scalar field. In principle, these can be obtained by acting on the scalar primary with all modes of the energy momentum tensor. Actually, we shall argue and check that, in the $n\to 1$ limit, it is enough
to consider descendants obtained by acting with $L_{-1}$ and $\overline L_{-1}$ alone, that is $\partial$, $\overline \partial$. This remark greatly simplifies the calculation and allowed us to check equality of (\ref{eq:CFTformula}) and (\ref{eq:Gravityformula}) for the terms with $t$ up to 5.

The terms with $r>1$ are more complicated and will be discussed later. On the gravity side they are associated with $r-$CDW group elements. On the CFT side, they come from multiple insertions of the scalar operator. 

\section{The gravity calculation}
\label{sec:gravity}

\subsection{Preliminary results}

Let us briefly review the results of \cite{Barrella:2013wja} in a compact way. Let $M$ be the $2\times 2$ 
diagonal matrix
\be
M = \mbox{diag}(e^{i\,\pi\,(1+1/n)}, \ e^{i\,\pi\,(1-1/n)}),
\ee
and $T$ the $2\times 2$ matrix
\be
T =\begin{pmatrix}
f(y) & f(-y) \\
-f(-y) & -f(y)
\end{pmatrix}
\ee
where $y$ is the branch of $x=\frac{4y}{(y+1)^{2}}$ such that $y=\frac{1}{4}x+\cdots$.
Remarkably, the matrix $T$ obeys the simple  property $y\,T^{2} = \mathbb I$. This means that 
\be
y\,[f(y)^{2}-f(-y)^{2}]=1,
\ee
and this relation fixes the even part of $T$ in terms of the odd part. In particular, from the $\mc O(y^{3})$
result of \cite{Barrella:2013wja}, we immediately obtain the $\mc O(y^{4})$ expression of $f(y)$
\ba
f(y) &=& \frac{n}{2 y}+\frac{1}{2 n}+\frac{\left(-n^4-n^2+2\right) y}{18 n^3}+\frac{\left(n^4+n^2-2\right) y^2}{18 n^5}\nonumber \\
&&
+\frac{\left(-76 n^8-80
   n^6+102 n^4+205 n^2-151\right) y^3}{4050 n^7}\nonumber \\
   &&+\frac{\left(101 n^8+130 n^6-177 n^4-305 n^2+251\right) y^4}{4050
   n^9}+O\left(y^{5}\right).
\ea
The Schottky generators are the matrices
\be
\mc L_{k} = M^{k}\,T^{-1}\,M\,T\,M^{-(k-1)},\qquad k=1, \dots, n-1.
\ee
At leading order $r=1$ in (\ref{eq:structure}), we need to consider the so-called consecutively decreasing words (CDW)
of the form 
\be
\gamma_{k, m} = \mc L_{k+m}\,\mc L_{k+m-1}\cdots\mc L_{m+1} = M^{m+k}\,T^{-1}\,M^{k}\,T\,M^{-m}.
\ee
The quantity $q_{\gamma}$ is defined as the root of  the secular equation
\be
\label{eq:secular}
\det(\gamma_{k,m}-q_{\gamma}^{-1/2}\,\mathbb I_{2\times 2})=0,
\ee
with $|q_{\gamma}|<1$. Clearly, it does not
depend on $m$, but is a function of $k$. Summing over the CDW elements, we have to remind that at each $k$ we have $n-k$ inequivalent
words, one for each $m$,  plus their inverses. Solving perturbatively (\ref{eq:secular}) and expanding in powers of $x$, we find 
\ba
q_{\gamma} &=& \frac{\kappa ^4 x^2}{16 n^4}+x^3 \left(\frac{\kappa ^6}{16 n^6}+\frac{\kappa ^4 \left(n^2-1\right)}{16 n^6}\right)+x^4 \left(\frac{7
   \kappa ^8}{128 n^8}+\frac{3 \kappa ^6 \left(n^2-1\right)}{32 n^8}+\frac{\kappa ^4 \left(n^2-1\right) \left(65 n^2-41\right)}{1152
   n^8}\right)\nonumber \\
   &&+x^5 \left(\frac{3 \kappa ^{10}}{64 n^{10}}+\frac{7 \kappa ^8 \left(n^2-1\right)}{64 n^{10}}+\frac{\kappa ^6
   \left(n^2-1\right) \left(83 n^2-59\right)}{768 n^{10}}+\frac{\kappa ^4 \left(n^2-1\right) \left(116 n^4-125 n^2+33\right)}{2304
   n^{10}}\right)\nonumber \\
   &&+x^6 \left(\frac{165 \kappa ^{12}}{4096 n^{12}}+\frac{15 \kappa ^{10} \left(n^2-1\right)}{128 n^{12}}+\frac{7 \kappa
   ^8 \left(n^2-1\right) \left(101 n^2-77\right)}{4608 n^{12}}+\frac{\kappa ^6 \left(n^2-1\right) \left(7 n^2-4\right) \left(25
   n^2-17\right)}{1536 n^{12}}\right.\nonumber \\
   &&\left. +\frac{\kappa ^4 \left(n^2-1\right) \left(374279 n^6-530301 n^4+244341 n^2-36479\right)}{8294400
   n^{12}}\right)\nonumber \\
   &&+x^7 \left(\frac{143 \kappa ^{14}}{4096 n^{14}}+\frac{495 \kappa ^{12} \left(n^2-1\right)}{4096 n^{14}}+\frac{5 \kappa
   ^{10} \left(n^2-1\right) \left(119 n^2-95\right)}{3072 n^{14}}\right. \nonumber\\
   &&\left. +\frac{7 \kappa ^8 \left(n^2-1\right) \left(246 n^4-337
   n^2+115\right)}{9216 n^{14}}+\frac{\kappa ^6 \left(n^2-1\right) \left(13298 n^6-22362 n^4+12417 n^2-2273\right)}{115200
   n^{14}}\right. \nonumber \\
   &&\left. +\frac{\kappa ^4 \left(n^2-1\right) \left(112679 n^8-190053 n^6+116829 n^4-30887 n^2+2952\right)}{2764800
   n^{14}}\right)+\mc O\left(x^8\right),
\ea
where we have defined 
\be
\kappa = \frac{1}{\sin\left(\frac{k\,\pi}{n}\right)}.
\ee

\subsection{The entanglement limit  $n=1$}

If we want to present results in the limit $n\to 1$, then several remarks simplify the calculation.
The quantity $q_{\gamma}$ involves even powers of $\kappa$ and is thus invariant under $k\to n-k$.
The contribution to the partition function of the various bulk fields can be expanded in powers of $x$, and the final summation over $k$ will always involve elementary sums of the form (the factor $2(n-k)$ counts the CDW)
\be
f_{n}(p) = \sum_{k=1}^{n-1}2\,(n-k)\,\frac{1}{\sin^{2p}\left(\frac{k\,\pi}{n}\right)} = n\,\sum_{k=1}^{n-1}\frac{1}{\sin^{2p}\left(\frac{k\,\pi}{n}\right)} 
\ee 
The analytic continuation of this sum has been discussed in \cite{CC2,Barrella:2013wja} and one finds the finite result
\be
\label{eq:basic}
\lim_{n\to 1}\frac{1}{n-1}f_{n}(p) = \frac{\sqrt{\pi }\, \Gamma (p+1)}{2\, \Gamma\left(p+\frac{3}{2}\right)}.
\ee
The fact that the singularity as $n\to 1$ is fully absorbed by the above sum has an important consequence. The $n\to 1$ limit
of the gravity partition function can be computed by setting from the scratch $n=1$ in the various coefficients of $q_{\gamma}$ keeping the dependence on $n$ only inside $\kappa$. Thus, we can work with the much simpler expression
\be
q_{\gamma} = \frac{\kappa ^4 x^2}{16}+\frac{\kappa ^6 x^3}{16}+\frac{7 \kappa ^8 x^4}{128}+\frac{3 \kappa ^{10} x^5}{64}+\frac{165 \kappa ^{12}
   x^6}{4096}+\frac{143 \kappa ^{14} x^7}{4096}+\mc O\left(x^8\right).
\ee
It is very tempting to resum this expression in the following closed form, at least perturbatively in $x$,
\be
\label{eq:resummation}
q_{\gamma} = \frac{4 \sqrt{1-\kappa ^2 x} \left(\kappa ^2 x-2\right)}{\kappa ^4 x^2}-\frac{8 \left(\kappa ^2 x-1\right)}{\kappa ^4 x^2}+1,
\ee
that means
\be
\label{eq:qgamma-explicit}
q_{\gamma} = 4\,\sum_{p=2}^{\infty}\bigg[
\binom{1/2}{p+1}+2\,\binom{1/2}{p+2}
\bigg]\,(-1)^{p+1}\,(\kappa^{2}\,x)^{p}.
\ee
This is an all order result for the CDW quantity $q_{\gamma}$ that can be used to obtain various explicit expansions.

\subsection{Results in the  higher spin current case}

The contribution we want to compute is that of a higher spin $s$ current
\be
S^{(s)\,\rm one-loop} = -\lim_{n\to 1}\frac{1}{n-1}\log\prod_{\gamma\in \mc P }\prod_{p=s}^{\infty}
\frac{1}{|1-q^{p}_{\gamma}|}.
\ee
We consider the CDW contribution that starts at $x^{2s}$ with corrections in powers of $x$. It is obtained by 
considering the subset $\mc P_{1}\subset \mc P$ of Schottky group elements that are CDW. For these elements, 
we have determined $q_{\gamma}$ in, say, (\ref{eq:qgamma-explicit}). We expand the logarithm
and find 
\ba
&& S^{(s)\, \rm one-loop}_{\rm CDW} = -\lim_{n\to 1}\frac{1}{n-1}\sum_{\gamma\in\mc P_{1}} \sum_{p=s}^{\infty}\bigg(q_{\gamma}^{p}+\frac{1}{2}q_{\gamma}^{2p}+\cdots\bigg) =  \\
&& = -\lim_{n\to 1}\frac{1}{n-1}\sum_{\gamma\in\mc P_{1}}\bigg(
\underbrace{\frac{q_{\gamma}^{s}}{1-q_{\gamma}}}_{(I)}+\underbrace{\frac{1}{2}\,\frac{q_{\gamma}^{2s}}{1-q_{\gamma}^{2}}}_{(II)}+\cdots\bigg)  = S^{(s)\, \rm one-loop}_{\rm CDW, (I)}+S^{(s)\, \rm one-loop}_{\rm CDW, (II)}+\cdots ,\nonumber
\ea
where we took intro account the fact that $q_{\gamma}$ is real. 
The first term (I) in the sum takes the form of $x^{2s}$ times a power series in $x$. Its terms 
up to $\mc O(x^{4s})$ are the exact contributions to the \reni entropy. The next terms must be combined with the 
second term (II) in the sum which takes the form of $x^{4s}$ times a power series in $x$. This part of the CDW 
contribution gets corrections from the \reni entropy due to the 2-CDW Schottky group elements that also starts at 
order $\mc O(x^{4s})$, see later sections.
Expanding $q_{\gamma}$ using 
(\ref{eq:resummation}) and performing the limit with (\ref{eq:basic}), we find the expansion
\ba
\label{eq:higher-result-I}
&& S^{(s)\, \rm one-loop}_{\rm CDW, (I)} =
-\left(\frac{x}{4}\right)^{2s}\,\frac{\sqrt\pi}{2}\,\frac{\Gamma(2s+1)}{\Gamma(2s+\frac{3}{2})}\,
\bigg[
1+\frac{2 s (2 s+1) x}{4 s+3}+\frac{(s+1) (2 s+1)^2 (4 s+1) x^2}{2 (4 s+3) (4
   s+5)}\nonumber\\
   &&+\frac{(s+1)^2 (2 s+1)^2 (2 s+3) x^3}{3 (4 s+5) (4 s+7)}+\frac{(s+1)^2 (s+2)
   (2 s+1) (2 s+3)^2 x^4}{12 (4 s+7) (4 s+9)}\nonumber\\
   &&+\frac{(s+1)^2 (s+2)^2 (2 s+1) (2 s+3)^2
   (2 s+5) x^5}{30 (4 s+3) (4 s+9) (4 s+11)}+\ordx{6}
\bigg],
\ea
and 
\ba
\label{eq:higher-result-II}
&& S^{(s)\, \rm one-loop}_{\rm CDW, (II)} =
-\left(\frac{x}{4}\right)^{4s}\,\frac{\sqrt\pi}{4}\,\frac{\Gamma(4s+1)}{\Gamma(4s+\frac{3}{2})}\,
\bigg[
1+\frac{4 s (4 s+1) x}{8 s+3}+\frac{2 s (2 s+1) (4 s+1) x^2}{8 s+5}\nonumber \\
&&+\frac{4 s
   (2 s+1)^2 (4 s+1) (4 s+3) x^3}{3 (8 s+3) (8 s+7)}+\frac{(s+1) (2 s+1)^2 (4
   s+1) (4 s+3) \left(256 s^3+448 s^2+204 s+3\right) x^4}{6 (8 s+3) (8 s+5) (8
   s+7) (8 s+9)}\nonumber \\
   &&+\frac{2 (s+1)^2 (2 s+1)^2 (4 s+1) (4 s+3) (4 s+5) \left(256
   s^3+576 s^2+348 s+15\right) x^5}{15 (8 s+3) (8 s+5) (8 s+7) (8 s+9) (8
   s+11)}+\mc O(x^6)
\bigg].
\ea
Of course, there is no difficulty in computing the analogous higher order terms $S^{(s)\, \rm one-loop}_{\rm CDW, (III)}$
and so on. The expansion  (\ref{eq:higher-result-I}) can be extended without difficulty, given the closed formula for 
$q_{\gamma}$. Doing so, one notices a remarkable regularity in its coefficients. After some work, we found
the following very nice compact formula
\be
S^{(s)\, \rm one-loop}_{\rm CDW, (I)} = 
-\frac{\sqrt\pi}{2}\,\left(\frac{x}{4}\right)^{2s}\,\sum_{p=0}^{\infty}
\frac{\Gamma(p+2s+1)\Gamma(2p+4s-1)}{\Gamma(p+1)\Gamma(p+2s+\frac{3}{2})\Gamma(p+4s-1)}\left(\frac{x}{4}\right)^{p}.
\ee
It can be summed in closed form with the very nice result
\be
S^{(s)\, \rm one-loop}_{\rm CDW, (I)} = -\frac{\sqrt\pi}{2}\,\left(\frac{x}{4}\right)^{2s}\,\frac{\Gamma(2s+1)}{\Gamma(2s+\frac{3}{2})}\,\,{}_{3}F_{2}\bigg(2s, 2s-\frac{1}{2}, 2s+1; 2s+\frac{3}{2}, 4s-1; x\bigg),
\ee
that we anticipated in the Introduction.

\subsubsection{Comparison with known results}

For generic $s$, the first two terms of (\ref{eq:higher-result-I}) agree with the general result of \cite{Perlmutter:2013paa}.
For the graviton, $s=2$, we obtain 
\ba
S^{(2)\,\rm one-loop}_{\rm CDW, (I)} &=& -\frac{x^4}{630}-\frac{2 x^5}{693}-\frac{15 x^6}{4004}-\frac{x^7}{234}-\frac{7 x^8}{1530}-\frac{84
   x^9}{17765}-\frac{x^{10}}{209}-\frac{10 x^{11}}{2093}+\mc O(x^{12}), \nonumber \\
S^{(2)\,\rm one-loop}_{\rm CDW, (II)} &=& -\frac{x^8}{437580}-\frac{2 x^9}{230945}-\frac{x^{10}}{51051}-\frac{10 x^{11}}{289731}+\mc O(x^{12}), \\
S^{(2)\,\rm one-loop}_{\rm CDW, (III)} &=& \mc O(x^{12}), \nonumber
\ea
and therefore the $\mc O(x^{12})$ CDW one-loop entropy is 
\ba
S^{(2)\,\rm one-loop}_{\rm CDW} &=& -\frac{x^4}{630}-\frac{2 x^5}{693}-\frac{15 x^6}{4004}-\frac{x^7}{234}-\frac{2003 x^8}{437580}-\frac{1094
   x^9}{230945}\nonumber \\
   && -\frac{4660 x^{10}}{969969}-\frac{9760 x^{11}}{2028117}+\mc O(x^{12}),
\ea
in full agreement with Eq.~(63) of \cite{Barrella:2013wja}.
For the spin $s=3$ current, we obtain 
\be
S^{(3)\,\rm one-loop}_{\rm CDW} = -\frac{x^6}{12012}-\frac{x^7}{4290}-\frac{7 x^8}{16830}-\frac{28 x^9}{46189}-\frac{15 x^{10}}{19019}-\frac{2
   x^{11}}{2093}+\mc O(x^{12}),
\ee
and the first three terms are in agreement with the $n\to 1$ limit of the $\mc O(x^{8})$ calculation (2.8) in \cite{Chen:2013dxa}. Finally, for the spin $s=4$ current, we obtain 
\be
S^{(4)\,\rm one-loop}_{\rm CDW} =-\frac{x^8}{218790}-\frac{4 x^9}{230945}-\frac{3
   x^{10}}{76076}-\frac{5 x^{11}}{71162}+\mc O(x^{12}),
\ee
and the first term is in agreement with the entanglement $n\to 1$ limit of the $\mc O(x^{8})$ calculation (2.9) in \cite{Chen:2013dxa}.

\subsection{Results in the  scalar field case}

In this case, the contribution we want to compute is that associated with (\ref{eq:detscalar})
\be
\label{eq:first-order}
S^{\rm (scalar)\,one-loop} = -\frac{1}{2}\,\lim_{n\to 1}\frac{1}{n-1}\log\prod_{\gamma\in \mc P }
\prod_{\ell,\ell'=0}^{\infty}
\frac{1}{1-q^{\ell+\frac{\Delta}{2}}_{\gamma}\,\overline q^{\ell'+\frac{\Delta}{2}}_{\gamma}}.
\ee
We can repeat the analysis of the previous section and write
\ba
S^{(s)\, \rm one-loop}_{\rm CDW} &=& -\frac{1}{2}\,\lim_{n\to 1}\frac{1}{n-1}\sum_{\gamma\in\mc P_{1}} \sum_{p=0}^{\infty}(p+1)\,\left(q_{\gamma}^{\Delta+p}+\frac{1}{2}q_{\gamma}^{2\,(\Delta+p)}+\cdots\right)\nonumber \\
&=& -\frac{1}{2}\,\lim_{n\to 1}\frac{1}{n-1}\sum_{\gamma\in\mc P_{1}} \left(
\underbrace{\frac{q_{\gamma}^{\Delta}}{(q_{\gamma}-1)^{2}}}_{(I)}+\underbrace{\frac{1}{2}\frac{q_{\gamma}^{2\,\Delta}}{(q_{\gamma}^{2}-1)^{2}}}_{(II)}+\cdots\right) = \nonumber \\
&&
= S^{\rm (scalar)\, \rm one-loop}_{\rm CDW, (I)}+S^{\rm (scalar)\, \rm one-loop}_{\rm CDW, (I)}+\cdots.
\ea
Expanding $q_{\gamma}$ using 
(\ref{eq:resummation}) and performing the limit with (\ref{eq:basic}), we find the expansions
\ba
\label{eq:scalar-result-I}
&& S^{\rm (scalar)\, one-loop}_{\rm CDW, (I)} =
-\left(\frac{x}{4}\right)^{2\Delta}\,\frac{\sqrt\pi}{4}\,\frac{\Gamma(2\Delta+1)}{\Gamma(2\Delta+\frac{3}{2})}\,
\bigg[\nonumber \\
&&
1+\frac{2 \Delta  (2 \Delta +1) x}{4 \Delta +3}+\frac{(\Delta +1) (2 \Delta +1) \left(4
   \Delta ^2+3 \Delta +1\right) x^2}{(4 \Delta +3) (4 \Delta +5)}\nonumber \\
   && +\frac{2 (\Delta +1)^2
   (2 \Delta +1) (2 \Delta +3) \left(4 \Delta ^2+5 \Delta +3\right) x^3}{3 (4 \Delta +3)
   (4 \Delta +5) (4 \Delta +7)}\nonumber \\
   &&+\frac{(\Delta +1) (\Delta +2) (2 \Delta +1) (2 \Delta
   +3) \left(32 \Delta ^4+144 \Delta ^3+262 \Delta ^2+237 \Delta +93\right) x^4}{12 (4
   \Delta +3) (4 \Delta +5) (4 \Delta +7) (4 \Delta +9)}\nonumber \\
   &&+\frac{(\Delta +1) (\Delta +2)^2
   (2 \Delta +1) (2 \Delta +3) (2 \Delta +5) \left(32 \Delta ^4+176 \Delta ^3+398 \Delta
   ^2+449 \Delta +225\right) x^5}{30 (4 \Delta +3) (4 \Delta +5) (4 \Delta +7) (4 \Delta
   +9) (4 \Delta +11)}+\mc O(x^6)\bigg],\nonumber
\ea
and
\ba
\label{eq:scalar-result-II}
&& S^{\rm (scalar)\, one-loop}_{\rm CDW, (II)} =
-\left(\frac{x}{4}\right)^{4\Delta}\,\frac{\sqrt\pi}{8}\,\frac{\Gamma(4\Delta+1)}{\Gamma(4\Delta+\frac{3}{2})}\,
\bigg[\nonumber \\
&&
1+\frac{4 \Delta  (4 \Delta +1) x}{8 \Delta +3}+\frac{2 \Delta  (2 \Delta +1) (4 \Delta +1) x^2}{8 \Delta
   +5}+\frac{4 \Delta  (2 \Delta +1)^2 (4 \Delta +1) (4 \Delta +3) x^3}{3 (8 \Delta +3) (8 \Delta
   +7)}\nonumber\\
   &&+\frac{(\Delta +1) (2 \Delta +1) (4 \Delta +1) (4 \Delta +3) \left(256 \Delta ^4+576 \Delta ^3+428
   \Delta ^2+105 \Delta +3\right) x^4}{3 (8 \Delta +3) (8 \Delta +5) (8 \Delta +7) (8 \Delta +9)}\nonumber\\
   &&+\frac{4
   (\Delta +1)^2 (2 \Delta +1) (4 \Delta +1) (4 \Delta +3) (4 \Delta +5) (8 \Delta +1) \left(32 \Delta ^3+84
   \Delta ^2+69 \Delta +15\right) x^5}{15 (8 \Delta +3) (8 \Delta +5) (8 \Delta +7) (8 \Delta +9) (8 \Delta
   +11)}\nonumber \\
   &&+\mc O(x^6).
\bigg]
\ea
The leading term is one half of the analogous result for the higher spin currents. This is due to the fact that 
the scalar partition function starts with the single state $|\frac{\Delta}{2},\frac{\Delta}{2}\rangle$, while the higher spin 
current provides two low-lying states $J_{-s}|0\rangle$ and $\overline J_{-s}|0\rangle$. In other words, the difference
is in the value of the multiplicity $\mc N$ appearing in (41) of \cite{Barrella:2013wja}. 
The bulk scalar field has been studied in the one-interval case in \cite{Datta:2013hba}.
No results are available in the two-interval case.

\subsection{Exponentiation in the classical limit}

It is interesting to take the limit of  (\ref{eq:higher-result-I}), (\ref{eq:higher-result-II}) , (\ref{eq:scalar-result-I}),
and (\ref{eq:scalar-result-II})
when $\Delta\to \infty$
with fixed product $\xi=s\,x$ (higher spin current) or $\xi=\Delta\,x$ (scalar). In the three cases, the square brackets 
takes the form of a simple exponential 
\be
\bigg[\cdots\bigg]_{(\ref{eq:higher-result-I})}\to e^{\xi}, \quad
\bigg[\cdots\bigg]_{(\ref{eq:higher-result-II})}\to e^{2\,\xi}, \quad
\bigg[\cdots\bigg]_{(\ref{eq:scalar-result-I})}\to e^{\xi}, \quad
\bigg[\cdots\bigg]_{(\ref{eq:scalar-result-II})}\to e^{2\,\xi}.
\ee
It would be very interesting to understand the relation between this exponentiation and the analogous properties of
the semiclassical conformal blocks \cite{Harlow:2011ny}.

\section{CFT check of the scalar field entanglement entropy}
\label{sec:CFT}

It is interesting to reproduce the result (\ref{eq:scalar-result-I}) from a direct CFT calculation according to (\ref{eq:CFTformula}).
To this aim, we consider a primary field $X(z, \overline z)$ with conformal weights $(L_{0}, \overline L_{0}) = (h, \overline h)$. In the end, we shall specialise to $h=\overline h = \frac{\Delta}{2}$.

The recipe (\ref{eq:CFTformula}) requires to list all quasi-primaries belonging to the replica space $(\mathbb C/\mb Z)^{\otimes n}$. If we are interested in reproducing (\ref{eq:scalar-result-I}), 
then we can restrict ourselves to linear combinations of operators of the form 
\be
\mc O(z, \overline z) = \mc A(z_{j_{1}}, \overline z_{j_{1}})\otimes \mc B(z_{j_{2}}, \overline z_{j_{2}}),
\ee
where $z_{j_{p}}$ is the image of $z$ on sheet $j_{p}$, and $j_{1}\neq j_{2}$. Also, $\mc A$ and $\mc B$ 
have to be linear in $X$. This means that $\mc A$, $\mc B$ are obtained acting on $X$ with the negative modes $L_{-n}$
of the energy momentum tensor $T$ (and its antiholomorphic counterpart). In principle, there are non trivial cases
involving $L_{-n}$ with $n>1$. For instance, the operator $\mc O = \mc A\otimes X$ with~\footnote{The operator
$(TX)$ is the conformal normal ordered product.}
\be
\mc A = (TX)-\frac{3}{2(2h+1)}\partial^{2}X,
\ee
is quasi-primary.
Nevertheless, in the $n\to 1$ limit, one can exploit a major simplification and consider only operators $\mc A, \mc B$
that are obtained from $X$ by applying a certain number of derivatives $\partial, \overline\partial$, forgetting at all $T$.

The reason behind this expectation is the following. The classical $\mc O(c)$ contribution to the entanglement  entropy
is expected to be universal. It is determined by the central charge only and fully encoded in the vacuum Verma module.
This has been checked in several examples \cite{Chen:2013kpa,Chen:2013dxa,Chen:2014kja}.
Any operator involving $T$ has a norm $\alpha_{K}$ growing with $c$. The associated coefficient $d_{K}$
can be expanded in inverse powers of the central charge $c$ and the leading term for any operator involving $X$ 
has to be accompanied by an explicit factor $(n^{2}-1)$. This has the consequence that the one-loop $\mc O(c^{0})$
contributions from such operators has the same factor making it vanish at $n=1$, since the sum over $j_{1}\neq j_{2}$
provides always another factor $n-1$ cancelling the $1/(n-1)$ in the definition of the \reni entropy. 

From the gravity point of view, 
the determinant formula (\ref{eq:detscalar}) counts precisely the descendants of $X$ obtained by acting with 
the Virasoro operators $L_{-1}, \overline L_{-1}$. The contributions from higher modes $L_{-n}$, $n>1$
is encoded in the one-loop partition function of the graviton bulk field, which is (\ref{eq:dethigher}) with $s=2$.

In the following part of this section, we shall construct all quasi-primaries of the above form and compute for all of them the normalisation $\alpha_{K}$ and the one-point coefficient $d_{K}$.

\subsection{Relevant quasi-primaries up to level 5}

We  define the level as $L_{0}+\overline L_{0}-2\,\Delta$, with $\Delta = h+\overline h$ being the scaling dimension of $X$. At level 0, there is a single field 
\ba
\mc O^{(0,1)} &=& XX, \\
\alpha^{(0,1)} &=& \alpha_{X}^{2}, \\
d^{(0,1)} &=&  \frac{1}{(2\,n\,s_{j_{1}j_{2}})^{2\Delta}}\,\alpha_{X},
\ea
where $s_{j_{1}j_{2}}=\sin\left(\pi\frac{j_{1}-j_{2}}{n}\right)$.
The notation is of course $XX\equiv X_{j_{1}}\otimes X_{j_{2}}$, and $\alpha_{X}^{2}$ is the normalisation of 
the 2-point function $\langle XX\rangle$. 

At level 1, we find the two conjugate operators
\ba
-i\,\mc O^{(1,1)} &=& \der X\,  X-X\,\der X, \\
\alpha^{(1,1)} &=& 4\,h\,\alpha_{X}^{2}, \\ 
d^{(1,1)} &=& 4\,h\,\frac{c_{j_{1}j_{2}}}{(2\,n\,s_{j_{1}j_{2}})^{2\Delta+1}}\,\alpha_{X},\\
\nonumber \\
-i\,\mc O^{(1,2)} &=& \derbar X\,  X-X\,\derbar X, \\
\alpha^{(1,2)} &=& 4\,\overline h\,\alpha_{X}^{2}, \\
d^{(1,2)} &=& 4\,\overline h\,\frac{c_{j_{1}j_{2}}}{(2\,n\,s_{j_{1}j_{2}})^{2\Delta+1}}\,\alpha_{X},
\ea
where $c_{j_{1}j_{2}}=\cos\left(\pi\frac{j_{1}-j_{2}}{n}\right)$.
In the following, it will be convenient to write $\overline {\mc O}$ to denote the operators that is obtained from $\mc O$
by swapping $\partial\leftrightarrow \overline\partial$. The coefficients $\alpha_{K}$ (as well as $d_{K}$) of the two operators are related by the swap $h\leftrightarrow \overline h$. Such pairs of conjugate operators give the same contribution to 
(\ref{eq:scalar-result-I}) which is evaluated at $h=\overline h$.

At level 2, the quasi-primaries are 
%
\ba
\mc O^{(2,1)} &=& \der X \, \der X-\frac{h}{2h+1}(\der^{2}X\, X+X\, \der^{2}X), \\
\alpha^{(2,1)} &=& \frac{4\,h^{2}(4h+1)}{2h+1}\,\alpha_{X}^{2}, \\ 
d^{(2,1)} &=& 2\,h\,\frac{4h+1-4h\,s_{j_{1}j_{2}}^{2}}{(2\,n\,s_{j_{1}j_{2}})^{2\Delta+2}}\,\alpha_{X},\\
\nonumber \\
\mc O^{(2,2)} &=& \overline{\mc O^{(2,1)}}, \\
\nonumber \\
\mc O^{(2,3)} &=& X\,\der\derbar X+\der\derbar X\,X-\der X\,\derbar X-\derbar X\,\der X, \\
\alpha^{(2,3)} &=& 16\,h\,\overline h\,\alpha_{X}^{2}, \\
d^{(2,3)} &=& 16\,h\,\overline h\,\frac{c_{j_{1}j_{2}}^{2}}{(2\,n\,s_{j_{1}j_{2}})^{2\Delta+2}}\,\alpha_{X}.
\ea
These operators are a subset of the operators considered in section (3.3) of \cite{Chen:2014kja}. As we mentioned before, 
we can skip any operator involving the energy-momentum tensor in this calculation. The next operators at levels 3, 4, 5
are 

{\bf Level 3}
%
\ba
-i\,\mc O^{(3,1)} &=& \der X\,\der\derbar X-\der\derbar X\,\der X+\frac{h}{2h+1}(\der^{2}\derbar X\,X-X\,\der^{2}\derbar X+\derbar X\,\der^{2}X-\der^{2}X\,\derbar X), \\
\alpha^{(3,1)} &=& \frac{16 h^{2}\overline h(4h+1)}{2h+1}\,\alpha_{X}^{2}, \\
d^{(3,1)} &=&  8\,h\,\overline h\,\frac{c_{j_{1}j_{2}}
\,(4h+1-4h\,s^{2}_{j_{1}j_{2}})}{(2\,n\,s_{j_{1}j_{2}})^{2\Delta+3}}\,\alpha_{X}, \\
\nonumber \\
\mc O^{(3,2)} &=& \overline{\mc O^{(3,1)}}, \\
\nonumber \\
-i\,\mc O^{(3,3)} &=& \der^{2}X\,\der X-\der X\,\der^{2}X+\frac{h}{3(h+1)}(X\der^{3}\,X-\der^{3}X\,X), \\
\alpha^{(3,3)} &=& \frac{16h^{2}(2h+1)(4h+3)}{3(h+1)}\,\alpha_{X}^{2}, \\
d^{(3,3)} &=& \frac{8}{3}\,h\,(2h+1)\,\frac{c_{j_{1}j_{2}}
\,(4h+3-4h\,s^{2}_{j_{1}j_{2}})}{(2\,n\,s_{j_{1}j_{2}})^{2\Delta+3}}\,\alpha_{X}, \\
\nonumber \\
\mc O^{(3,4)} &=& \overline{\mc O^{(3,3)}}.
\ea

{\bf Level 4}
\ba
\mc O^{(4,1)} &=& \der^{3}X\,\der X+\der X\,\der^{3}X-\frac{h}{2(2h+3)}(
\der^{4}X\,X+X\,\der^{4}X)-\frac{3(h+1)}{2h+1}\,\der^{2}X\,\der^{2}X, \\
\alpha^{(4,1)} &=& \frac{48h^{2}(h+1)(4h+3)(4h+5)}{2h+3}, \\
d^{(4,1)} &=&  4\,h\,(h+1)\,\frac{16h^{2}c^{4}_{j_{1}j_{2}}+8h(s^{4}_{j_{1}j_{2}}-5s^{2}_{j_{1}j_{2}}+4)-12s^{2}_{j_{1}j_{2}}+15}{(2\,n\,s_{j_{1}j_{2}})^{2\Delta+4}}\,\alpha_{X}, \\
\nonumber \\
\mc O^{(4,2)} &=& \overline{\mc O^{(4,1)}}, \\
\nonumber \\
\mc O^{(4,3)} &=& \der^{2}X\,\der\derbar X+\der\derbar X\,\der^{2}X-\der X\,\der^{2}\derbar X-\der^{2}
\derbar X\,\der X\nonumber \\
&&+\frac{h}{3(h+1)}(\der^{3}\derbar X\,X-\der^{3}X\,\derbar X-\derbar X\,\der^{3}X+X\,
\der^{3}\derbar X), \\
\alpha^{(4,3)}&=&\frac{64 h^{2}\overline h (2h+1)(4h+3)}{3(h+1)}, \\
d^{(4,3)} &=& \frac{32}{3}\,h\,\overline h\,(2h+1)\,\frac{c^{2}_{j_{1}j_{2}}(4h\,c^{2}_{j_{1}j_{2}}+3)}
{(2\,n\,s_{j_{1}j_{2}})^{2\Delta+4}}\,\alpha_{X}, \\
\nonumber \\
\mc O^{(4,4)} &=& \overline{\mc O^{(4,3)}}, \\
\nonumber \\
\mc O^{(4,5)} &=& \der\derbar X\,\der\derbar X-\frac{h}{2h+1}(\der^{2}\derbar X\,\derbar X+\derbar X\,
\der^{2}\derbar X)
-\frac{\overline h}{2\overline h+1}(\derbar^{2}\der X\,\der X+\der X\,
\derbar^{2}\der X)\nonumber \\
&& +\frac{h\overline h}{(2h+1)(2\overline h+1)}(\der^{2}\derbar^{2}X\,X+\der^{2}X\,\derbar^{2}X
+\derbar^{2}X\,\der^{2}X+X\,\der^{2}\derbar^{2}X), \\
\alpha^{(4,5)} &=& \frac{16h^{2}\overline h^{2}(4h+1)(4\overline h+1)}{(2h+1)(2\overline h+1)}, \\
d^{(4,5)} &=& 4\,h\,\overline h\,\frac{(4h c^{2}_{j_{1}j_{2}}+1)(4\overline h c^{2}_{j_{1}j_{2}}+1)}{(2\,n\,s_{j_{1}j_{2}})^{2\Delta+4}}\,\alpha_{X}.
\ea

{\bf Level 5}

Let us define $\mk s = s_{j_{1}j_{2}}$, $\mk c=c_{j_{1}j_{2}}$.
%
\ba
-i \mc O^{(5,1)} &=& \der^{2}X\,\der^{3}X-\der^{3}X\,\der^{2}X+\frac{h(2h+1)}{10(h+2)(2h+3)}
(X\,\der^{5}X-\der^{5}X\,X)\nonumber \\
&&+\frac{2h+1}{2(2h+3)}(\der^{4}X\,\der X-\der X\,\der^{4}X), \\
\alpha^{(5,1)} &=& \frac{192 h^{2}(h+1)(2h+1)^{2}(4h+5)(4h+7)}{5(h+2)(2h+3)}, \\
d^{(5,1)} &=& \frac{16}{5}h(h+1)(2h+1)\frac{\mk c\,(16h^{2}\mk c^{4}+8h(\mk s^{4}-7\mk s^{2}+6)-20\mk s^{2}+35)}{(2\,n\,s_{j_{1}j_{2}})^{2\Delta+5}}\,\alpha_{X}, \\
\nonumber \\
\mc O^{(5,2)} &=& \overline{\mc O^{(5,1)}}, \\
\nonumber \\
-i\mc O^{(5,3)} &=&\der^{2}\derbar X\,\der^{2}X-\der^{2}X\,\der^{2}\derbar X\nonumber \\
&& +\frac{h(2h+1)}{6(h+1)(2h+3)}(\der^{4}\derbar X\,X-\der^{4}X\,\derbar X+\derbar X\,\der^{4}X-X\der^{4}
\derbar X)\nonumber\\
&&+\frac{2h+1}{3(h+1)}(\der^{3}X\,\der\derbar X-\der^{3}\derbar X\,\der X-\der\derbar X\,\der^{3}X+\der X\,\der^{3}
\derbar X), \\
\alpha^{(5,3)} &=& \frac{64 h^{2}\overline h (2h+1)^{2}(4h+3)(4h+5)}{3(h+1)(2h+3)}, \\
d^{(5,3)}&=& \frac{16}{3}h\,\overline h\,(2h+1)\frac{\mk c\,(16h^{2}\mk c^{4}+8h(\mk s^{4}-5\mk s^{2}+4)-12\mk s^{2}+15)}{(2\,n\,s_{j_{1}j_{2}})^{2\Delta+5}}\,\alpha_{X}, \\
\nonumber \\
\mc O^{(5,4)} &=& \overline{\mc O^{(5,3)}}, \\
\nonumber \\
-i\mc O^{(5,5)} &=& \der^{2}\derbar X\,\der\derbar X-\der\derbar X\,\der^{2}\derbar X+\frac{h}{3(h+1)}
(\derbar X\,\der^{3}\derbar X-\der^{3}\derbar X\,\derbar X)\nonumber \\
&&+\frac{\overline h}{2\overline h+1}(\der X\,\der^{2}\derbar^{2}X+\der\derbar^{2}X\,\der^{2}X
-\der^{2}X\,\der\derbar^{2}X-\der^{2}\derbar^{2}X\,\der X)\nonumber\\
&&+\frac{h\overline h}{3(h+1)(2\overline h+1)}(
\der^{3}\derbar^{2}X\,X+\der^{3}X\,\derbar^{2}X-\derbar^{2}X\,\der^{3}X-X\,\der^{3}\derbar^{2}X), \\
\alpha^{(5,5)}&=& \frac{64h^{2}\overline h^{2}(2h+1)(4h+3)(4\overline h+1)}{3(h+1)(2\overline h+1)}, \\
d^{(5,5)} &=& \frac{16}{3}h\,\overline h\,(2h+1)\frac{\mk c\,(4h \mk c^{2}+3)(4\overline h \mk c^{2}+1)}{(2\,n\,s_{j_{1}j_{2}})^{2\Delta+5}}\,\alpha_{X}, \\
\nonumber \\
\mc O^{(5,6)} &=& \overline{\mc O^{(5,5)}}.
\ea

\subsection{Matching the gravity side}

We now plug the results collected in the last section into 
(\ref{eq:CFTformula}) and expand the hypergeometric conformal blocks in order to recover the full  
$\mc O(x^{2\Delta+5})$ gravity result (\ref{eq:scalar-result-I}) after going to the $n\to 1$ limit. Notice that the 
involved sums are the same as in the gravity calculation since the sum over $j_{1}<j_{2}$ gives back the 
sum over $k=|j_{2}-j_{1}|$ with multiplicity $n-k$. The first two terms reproduce the calculation 
discussed in \cite{Perlmutter:2013paa}. Consider now, as an example, the next term which is at $\mc O(x^{2\Delta+2})$.
The relevant terms in (\ref{eq:CFTformula}) are
\ba
&&
-\frac{1}{2\,(2n)^{4\Delta}}\bigg\{
\frac{1}{s_{k}^{4\Delta}} \bigg[
1+2 h x+\frac{h \left(8 h^2+5 h+1\right) x^2}{4
   h+1}+\cdots
\bigg]\nonumber \\
&&+2\Delta\frac{1}{(2n)^{2}}\frac{c_{k}^{2}}{s_{k}^{4\Delta+2}}
\bigg[2+(4 h +1) x+\cdots\bigg]\,x \nonumber \\
&& + 
x^{2}\bigg(
\frac{16}{(2n)^{4}}h^{2}\frac{c_{k}^{4}}{s_{k}^{4\Delta+4}}+
\frac{2}{(2n)^{4}}\frac{2h+1}{4h+1}\frac{(4h+1-4h s_{k}^{2})^{2}}{s_{k}^{4\Delta+4}}
\bigg)\,x^{2}+\cdots
\bigg\}
\ea
where $s_{k} = \sin(\pi\,k/n)$, $c_{k}=\cos(\pi\,k/n)$, $h = \Delta/2$, and the square brackets are the expansions of the
conformal blocks. Going to the $n\to 1$ limit we reduce the coefficient of $x^{2}$ in this expression to the simple quantity
\be
-4^{-2\Delta-2}(4\Delta^{2}+3\Delta+1)\frac{1}{s_{k}^{4\Delta+4}}
\stackrel{(\ref{eq:basic})}{\longrightarrow}
-\frac{\sqrt{\pi }\, 2^{-4 \Delta -5} \left(4 \Delta ^2+3
   \Delta +1\right) \Gamma (2 \Delta +3)}{\Gamma \left(2
   \Delta +\frac{7}{2}\right)},
\ee
where the r.h.s. is the result of the sum over $k$ by means of  (\ref{eq:basic}). This is in full agreement with the 
$\mc O(x^{2\Delta+2})$ in  (\ref{eq:scalar-result-I}). A similar check can be repeated for the other terms of 
 (\ref{eq:scalar-result-I}) with full agreement.
 
 \section{The 2-CDW contributions}
 \label{sec:2-CDW}
 
 In this section, we discuss the 2-CDW contributions providing an exact formula for the leading $\ordx{4s}$ term
 and the tools needed to evaluate the next two terms $\ordx{4s+1}$ and $\ordx{4s+2}$. In particular, we shall combine
 these results with the analysis of the previous sections in order to compute the exact entanglement entropy for a spin $s=2,3,4$ current at order $\ordx{4s+2}$.

\subsection{Preliminary remarks}

As discussed in \cite{Barrella:2013wja}, the 2-CDW contributions are obtained by introducing the 
2-words Schottky group elements
\be
\gamma = \gamma_{[m_{1},m_{2}];[m_{3},m_{4}]} = M^{m_{1}}\,T^{-1}\,M^{m_{1}-m_{2}}\,T\,
M^{-m_{2}+m_{3}}\,T^{-1}\,M^{m_{3}-m_{4}}\,T\,M^{-m_{4}},
\ee
and computing the associated eigenvalue $q_{\gamma}^{-1/2}$ from (\ref{eq:secular}) . The sum over 2-CDW
words is  performed by summing over $\{m_{i}\}$ in the range
\be
0\le m_{i}\le n-1, \ m_{1}\neq m_{2},\ m_{3}\neq m_{4},\ m_{2}\neq m_{3},\ m_{4}\neq m_{1},\ 
(m_{1},m_{2})\neq (m_{3},m_{4}),
\ee
with a global over counting factor $1/2$. An explicit calculation gives the expansion
\ba
q_{\gamma} &=&   \left(\frac{x}{4}\right)^{4}\, \frac{1}{\mathfrak{s}_{m_1-m_2}^2 \mathfrak{s}_{m_2-m_3}^2 \mathfrak{s}_{m_1-m_4}^2 \mathfrak{s}_{m_3-m_4}^2}\nonumber \\
&& +\left(\frac{x}{4}\right)^{5}\,\frac{\mathfrak{c}_{2 m_1-2 m_2+2 m_3-2 m_4}+8 \mathfrak{s}_{m_1-m_2} \mathfrak{s}_{m_2-m_3} \mathfrak{s}_{m_1-m_4}
   \mathfrak{s}_{m_3-m_4}-1}{\mathfrak{s}_{m_1-m_2}^3 \mathfrak{s}_{m_2-m_3}^3 \mathfrak{s}_{m_1-m_4}^3 \mathfrak{s}_{m_3-m_4}^3}\nonumber \\
   &&+\left(\frac{x}{4}\right)^{6}\,\frac{1}{8 \mathfrak{s}_{m_1-m_2}^4 \mathfrak{s}_{m_2-m_3}^4 \mathfrak{s}_{m_1-m_4}^4
   \mathfrak{s}_{m_3-m_4}^4}
\bigg(
   3 \mathfrak{c}_{4 m_1-4 m_2}-32 \mathfrak{c}_{2 m_1-2 m_2}+3 \mathfrak{c}_{4 m_1-4 m_3}+3 \mathfrak{c}_{4 m_2-4 m_3}\nonumber \\
   &&+32
   \mathfrak{c}_{2 m_1-2 m_3}-6 \mathfrak{c}_{2 \left(m_1+m_2-2 m_3\right)}-32 \mathfrak{c}_{2 m_2-2 m_3}+6 \mathfrak{c}_{2 \left(m_1-2
   m_2+m_3\right)}-6 \mathfrak{c}_{4 m_1-2 \left(m_2+m_3\right)}+3 \mathfrak{c}_{4 m_1-4 m_4}\nonumber \\
   &&+3 \mathfrak{c}_{4 m_2-4 m_4} 
   +3
   \mathfrak{c}_{4 m_3-4 m_4}-32 \mathfrak{c}_{2 m_1-2 m_4}-6 \mathfrak{c}_{2 \left(m_1+m_2-2 m_4\right)}+32 \mathfrak{c}_{2 m_2-2
   m_4}+6 \mathfrak{c}_{2 \left(m_1+m_3-2 m_4\right)}\nonumber \\
   &&-6 \mathfrak{c}_{2 \left(m_2+m_3-2 m_4\right)}-32 \mathfrak{c}_{2 m_3-2 m_4}+12
   \mathfrak{c}_{2 \left(m_1+m_2-m_3-m_4\right)}+8 \mathfrak{c}_{2 \left(m_1-m_2+m_3-m_4\right)}-6 \mathfrak{c}_{2 \left(m_1-2
   m_2+m_4\right)}\nonumber \\
   &&-6 \mathfrak{c}_{2 \left(m_1-2 m_3+m_4\right)}+6 \mathfrak{c}_{2 \left(m_2-2 m_3+m_4\right)}+12 \mathfrak{c}_{2
   \left(m_1-m_2-m_3+m_4\right)}+6 \mathfrak{c}_{4 m_1-2 \left(m_2+m_4\right)}-6 \mathfrak{c}_{4 m_1-2 \left(m_3+m_4\right)}\nonumber \\
   &&-6
   \mathfrak{c}_{4 m_2-2 \left(m_3+m_4\right)}+38\bigg)+\mc O(x^{7}),
   \ea
   where
   \be
   \mk s_{m} = \sin\frac{\pi\,m}{n},\qquad
   \mk c_{m} = \cos\frac{\pi\,m}{n}.
   \ee
Unfortunately, the structure of  the dependence on the four integers $m_{i}$ is now quite complicated and no simple regularity is observed in the higher terms. Hence, we cannot provide an all-order expression of $q_{\gamma}$ neither in the $n\to 1$ limit. 
   
\section{The leading 2-CDW term}

The leading contribution to the entanglement entropy is 
\be
S^{(s)\,\rm one-loop}_{\rm 2-CDW} = -\lim_{n\to 1}\frac{1}{n-1}\,\frac{1}{2}\sum_{\mc P_{2}}\bigg(
\frac{q_{\gamma}^{s}}{1-q_{\gamma}}+\cdots\bigg),
\ee
where $\mc P_{2}$ denotes the sum over $m_{i}$ in the above range, and the dots inside the round bracket 
give a contribution that starts at order $x^{8s}$. From the expansion of $q_{\gamma}$, we obtain 
\be
\label{eq:cdw2-formula}
S^{(s)\,\rm one-loop}_{\rm 2-CDW} =  
-\lim_{n\to 1}\frac{1}{n-1}\,\frac{1}{2}\,\left(
\frac{x}{4}\right)^{4s}\,F_{s}(n) +\mc O(x^{4s+1}),
\ee
where
\be
F_{s}(n) = \sum_{\mc P_{2}}
\frac{1}{\sin ^{2s}\left(\frac{\pi  \left(m_1-m_2\right)}{n}\right) \sin
   ^{2s}\left(\frac{\pi  \left(m_2-m_3\right)}{n}\right) \sin ^{2s}\left(\frac{\pi 
   \left(m_1-m_4\right)}{n}\right) \sin ^{2s}\left(\frac{\pi 
   \left(m_3-m_4\right)}{n}\right)}.
   \ee
We want to obtain an explicit formula for this contribution for a generic (integer $\ge 2$) value of $s$. To this aim, we 
begin by remarking that $F_{s}(n)$ can be written as
\be
F_{s}(n) = \mbox{Tr}(H_{n}(s)^{4})-\mbox{Tr}(H_{n}(2s)^{2}),
\ee
where $H_{n}(s)$ is the $n\times n$ matrix
\be
\left[H_{n}(s)\right]_{pq} = \left\{\begin{array}{ll}
\csc^{2s}\left(\frac{p-q}{n}\right),& \quad p\neq q, \\
0, & \quad p=q.
\end{array}\right. 
\ee
This matrix is circulant and therefore its eigenvalues are known and read
\be
\lambda^{(p)}_{n}(s) = \sum_{k=1}^{n-1}\cos\left(\frac{2\,\pi\,p\,k}{n}\right)\frac{1}{\sin^{2s}\left(\frac{\pi\,k}{n}\right)}, \qquad p =0, \dots, n-1.
\ee
Thus, 
\be
F_{s}(n) = \sum_{p=0}^{n-1}\left[\lambda^{(p)}_{n}(s)\right]^{4}-\sum_{p=0}^{n-1}\left[\lambda^{(p)}_{n}(2s)\right]^{2}.
\ee
At fixed integer $s$, the eigenvalues $\lambda^{(p)}_{n}(s)$ are polynomials in both $p$ and $n$. We checked that the 
$n\to 1$ limit of $1/(n-1)$ times $F_{s}(n)$ can be computed by setting $n=1$ in those polynomials. The cancelling 
factor $n-1$ comes entirely from the sum over $p$. Hence, inside $F_{s}(n)$, we can exploit the analytically continued expression
\be
\lambda_{n\to 1}^{(p)}(s) = -\frac{(-4)^{s}}{\Gamma(2s+1)}(p-s)_{2s},
\ee
where $(a)_{b}$ is the Pochammer symbol. This is a finite polynomial and the final step requires the following
formula
\be
\lim_{n\to 1}\frac{1}{n-1}\sum_{p=0}^{n-1}p^{N} = -N\,\zeta(1-N).
\ee
In conclusion, 
\be
S^{(s)\,\rm one-loop}_{\rm 2-CDW} = \sigma_{s}\,x^{4s}+\mc O(x^{4s+1}), 
\ee
where the coefficient $\sigma_{s}$ is obtained from 
\be
\label{eq:sigma}
\sigma_{s} =  -\frac{1}{2}\left[
\frac{1}{\Gamma(2s+1)^{4}}\,(p-s)_{2s}^{4}-\frac{1}{\Gamma(4s+1)^{2}}\,(p-2s)_{4s}^{2}
\right]_{p^{N}\to -N\,\zeta(1-N)},
\ee
as mentioned in the Introduction. 
Just to give a check of this formula, let us consider the three cases $s=2,3,4$. For $s=2$, we find
\ba
F_{2}(n) &=& \frac{4n \left(n^2-4\right) \left(n^2-1\right)}{488462349375}\, (5703 n^{12}+192735 n^{10}+3812146 n^8+75493430
   n^6+1249638099 n^4\nonumber \\
   &&+9895897835 n^2-162763727948),
\ea
in agreement with \cite{Barrella:2013wja}. The evaluation of  (\ref{eq:cdw2-formula}) agrees with the coefficient (\ref{eq:sigma}). For $s=3$ and $s=4$, we find
\ba
F_{3}(n) &=& \frac{2n \left(n^2-4\right) \left(n^2-1\right)}{3028793579456347828125}\,(7121028183 n^{20}+318929181915
   n^{18}+7383843717075 n^{16}\nonumber \\
   && +120385864634215 n^{14}+1646919142515245 n^{12}+23516641325405865 n^{10}\nonumber \\
   && +379716725855421985
   n^8+6695285026854147965 n^6+98177254369909711080 n^4\nonumber \\
   &&+509870981187643230040 n^2-18438453061190079393568), \\
F_{4}(n) &=& \frac{4n \left(n^2-4\right) \left(n^2-1\right)}{7217620475953080409439269921875}\,(890779990098075
   n^{28}+51377796296763575 n^{26}\nonumber \\
   &&+1506538876909209643 n^{24}+30021482057949152715 n^{22}+460566249653162829602
   n^{20}\nonumber\\
   &&+5915955950696734741150 n^{18}+69830671794873569438670 n^{16}\nonumber \\
   &&+879907717975387768404750
   n^{14}+13575412535055352666915815 n^{12}\nonumber \\
&&   +253558442419836251074578075 n^{10}+4969225821536517978217020295
   n^8\nonumber \\
   &&+87557592432594348779328719975 n^6+1059813722330022285316966964892 n^4 \\
   &&+3656969159093966862988863639760
   n^2-232004722704441042012231546476992).\nonumber 
\ea
and, again, (\ref{eq:cdw2-formula}) agrees  (\ref{eq:sigma}). Clearly, the formula sigma is a very convenient way of computing $\sigma_{s}$ at any $s$ since its evaluation is almost instantaneous on any algebraic manipulation system, even for large $s$.

If one is interested in the entanglement entropy of a scalar field at a certain $\Delta$, the calculation is similar
and the leading contribution is easily obtained from the formula (\ref{eq:sigma}). A CFT calculation would require the 
analysis of operators quartic in the scalar operator $X(z, \overline z)$.

\section{The complete entanglement entropy of a spin $s=2,3,4$ current at $\ordx{4s+2}$}
\label{sec:complete}

To summarise, we present the first three 2-CDW contributions to the cases $s=2,3,4$. The leading term is computed using 
(\ref{eq:sigma}). The other two are determined by evaluating the sums over $\mc P_{2}$ as specific polynomials in $n$ and taking the $n\to 1$ limit in the definition of the entanglement entropy. The result is 
\ba
S^{(s=2)\,\rm one-loop}_{\rm 2-CDW} &=&-\frac{29 x^8}{510510}-\frac{100 x^9}{2909907}-\frac{14 x^{10}}{138567}+\ordx{11},\\
S^{(s=3)\,\rm one-loop}_{\rm 2-CDW} &=&-\frac{313 x^{12}}{148728580}-\frac{392 x^{13}}{239028075}-\frac{29321 x^{14}}{4621209450}+\ordx{15}, \\
S^{(s=4)\,\rm one-loop}_{\rm 2-CDW} &=&-\frac{113393 x^{16}}{1289317436550}-\frac{648 x^{17}}{8796512725}-\frac{444896 x^{18}}{1236789689135}+\ordx{19}.
\ea
These contributions must be added to the $S^{(s)\,\rm one-loop}_{\rm CDW, (I)}$  and $S^{(s)\,\rm one-loop}_{\rm CDW, (II)}$ terms that we computed previously and that read
\ba
S^{(s=2)\,\rm one-loop}_{\rm CDW, (I)} &=&-\frac{x^4}{630}-\frac{2 x^5}{693}-\frac{15 x^6}{4004}-\frac{x^7}{234}-\frac{7 x^8}{1530}-\frac{84
   x^9}{17765}-\frac{x^{10}}{209}+\ordx{11}, \\
S^{(s=2)\,\rm one-loop}_{\rm CDW, (II)} &=&-\frac{x^8}{437580}-\frac{2 x^9}{230945}-\frac{x^{10}}{51051}
+\ordx{11},\\
S^{(s=3)\,\rm one-loop}_{\rm CDW, (I)} &=&-\frac{x^6}{12012}-\frac{x^7}{4290}-\frac{7 x^8}{16830}-\frac{28 x^9}{46189}-\frac{15 x^{10}}{19019}-\frac{2
   x^{11}}{2093}\nonumber \\
   &&-\frac{33 x^{12}}{29900}-\frac{11 x^{13}}{8925}-\frac{143 x^{14}}{106488}+\ordx{15}, \\
S^{(s=3)\,\rm one-loop}_{\rm CDW, (II)} &=&-\frac{x^{12}}{135207800}-\frac{x^{13}}{23401350}-\frac{3 x^{14}}{21544100}+\ordx{15}, \\
S^{(s=4)\,\rm one-loop}_{\rm CDW, (I)} &=&-\frac{x^8}{218790}-\frac{4 x^9}{230945}-\frac{3 x^{10}}{76076}-\frac{5 x^{11}}{71162}-\frac{11 x^{12}}{101660}-\frac{11
   x^{13}}{72675}-\frac{1001 x^{14}}{5058180}\nonumber \\
   && -\frac{143 x^{15}}{580754}-\frac{65 x^{16}}{220286}-\frac{104
   x^{17}}{302841}-\frac{221 x^{18}}{565915}+\ordx{19}, \\
S^{(s=4)\,\rm one-loop}_{\rm CDW, (II)} &=&-\frac{x^{16}}{39671305740}-\frac{4 x^{17}}{20419054425}-\frac{2 x^{18}}{2398428615}+\ordx{19}.
\ea
In conclusion, the one-loop entanglement entropies for $s=2,3,4$ at order $\ordx{4s+2}$ are
\ba
S^{(s=2)\,\rm one-loop} &=& -\frac{x^4}{630}-\frac{2 x^5}{693}-\frac{15 x^6}{4004}-\frac{x^7}{234}-\frac{167 x^8}{36036}-\frac{69422
   x^9}{14549535}-\frac{122 x^{10}}{24871}+\ordx{11}, \\
S^{(s=3)\,\rm one-loop} &=&-\frac{x^6}{12012}-\frac{x^7}{4290}-\frac{7 x^8}{16830}-\frac{28 x^9}{46189}-\frac{15 x^{10}}{19019}-\frac{2
   x^{11}}{2093}-\frac{1644627 x^{12}}{1487285800}\nonumber \\
   && -\frac{458893 x^{13}}{371821450}-\frac{224484047
   x^{14}}{166363540200}+\ordx{15}, \\
S^{(s=4)\,\rm one-loop} &=&-\frac{x^8}{218790}-\frac{4 x^9}{230945}-\frac{3 x^{10}}{76076}-\frac{5 x^{11}}{71162}-\frac{11
   x^{12}}{101660}\nonumber \\
   &&-\frac{11 x^{13}}{72675}-\frac{1001 x^{14}}{5058180}
    -\frac{143 x^{15}}{580754}-\frac{253702367
   x^{16}}{859544957700}\nonumber \\
   && -\frac{1550029508 x^{17}}{4512611027925}-\frac{207186247
   x^{18}}{530052723915}+\ordx{19},
\ea
which are the results summarised in the Introduction.

\section{Conclusions}

In this paper, we have studied the  one-loop correction to the holographic entanglement entropy in $AdS_{3}/CFT_{2}$.
We have considered the specific contribution due to a bulk higher spin $s$ current or a scalar field 
with full scaling dimension $\Delta$. The case of the scalar field is an important ingredient of Vasiliev higher spin 
gravity theory holographically dual to large $N$ conformal coset models. In all cases, we focused on the 
non-trivial two-interval case at small cross ratio $x$ where a perturbative analysis can be performed. 
Our main concern has been the $n\to 1$ entanglement limit of the $n$-th order \reni entropy. In this limit, 
various simplifications occur already on the gravity side of AdS/CFT. In particular, the CDW contribution to the
entanglement entropy can be obtained in closed form for the higher spin current at least 
for the contributions of the form $\mc O(x^{2s+p})$ up to $\mc O(x^{4s})$, where 2-CDW elements have to be
considered. In the scalar field case, we have computed the similar contributions for generic values of $\Delta$. 
As a check of the calculation, we have compared the terms up to $\mc O(x^{2\Delta+5})$  with an explicit CFT calculation finding full agreement. This is clearly not surprising, but the calculation is nevertheless interesting.
Indeed, in the entanglement limit, there are simplifications on the CFT side too. 
In particular, one can check that the set of relevant operators is quite restricted. It is not necessary to take into 
account the quasi-primaries built with the energy-momentum tensor. This is related to the important fact that scalars
do not alter the leading semiclassical entropy that depends universally only on the central charge. 
At the next-to-leading one-loop order, the contributions from the various bulk fields are decoupled this is the reason 
behind the simplicity of our CFT comparison. We have also analysed the corrections to the entanglement entropy
from the 2-CDW contributions. In particular, we have found an explicit formula for the leading $\ordx{4s}$
correction as a function of $s$.

%

\providecommand{\href}[2]{#2}\begingroup\raggedright\endgroup

\end{document}